\documentclass[aps,twocolumn]{revtex4}
\usepackage[colorlinks=true, pdfstartview=FitV, linkcolor=blue, citecolor=red, urlcolor=magenta, breaklinks=true]{hyperref}
\usepackage{color}
\usepackage{subeqnarray}
\usepackage{graphicx}
\usepackage[all]{xy}
\usepackage{amsmath}
\usepackage{amssymb}
\usepackage{subfigure}
\usepackage{mwe}
\newcommand{\be}{\begin{equation}}
\newcommand{\ee}{\end{equation}}
\def\bal#1\eal{\begin{align}#1\end{align}}
\newcommand{\ben}{\begin{eqnarray}}
\newcommand{\een}{\end{eqnarray}}
\newcommand{\bes}{\begin{subequations}}
\newcommand{\ees}{\end{subequations}}
\newcommand{\bens}{\begin{subeqnarray}}
\newcommand{\eens}{\end{subeqnarray}}

\newcommand{\bb}{\bibitem}
\def\tanh{\text{tanh}}

\def\Ci{\text{Ci}}

\begin{document}
\title{Symmetric and asymmetric thick brane structures}
\author{D. Bazeia}
\affiliation{Departamento de F\'\i sica, Universidade Federal da Para\'\i ba, 58051-970 Jo\~ao Pessoa, PB, Brazil}

\author{D.A. Ferreira}
\affiliation{Departamento de F\'\i sica, Universidade Federal da Para\'\i ba, 58051-970 Jo\~ao Pessoa, PB, Brazil}

\author{M.A. Marques}
\affiliation{Departamento de F\'\i sica, Universidade Federal da Para\'\i ba, 58051-970 Jo\~ao Pessoa, PB, Brazil}

\vspace{3cm}
\date{\today}

\begin{abstract}

This work deals with the presence of thick branes in a model with two source scalar fields that interact with one another in a very specific way. The model is new, capable of generating kinklike configurations that engender important modifications in the energy density of the system, inducing the presence of symmetric and asymmetric structures inside the brane.\\

\end{abstract}


\maketitle

{\bf 1. Introduction.} The braneworld scenario first proposed in \cite{Ra} in the form of a thin brane, was soon generalised to describe thick branes in the presence of scalar fields; see, e.g., Refs. \cite{gw,De,thick1,thick2,thick3,thick4,bbn,thick5,bloch,branestructure,melfo,dutra,gomes} and references therein. In the thick brane scenario the models deal with a five-dimensional $AdS$ warped geometry with a single extra dimension of infinite extent in the presence of real scalar fields as source fields to generate the brane. For a review see Ref. \cite{book}.

Consistency of superstring theory \cite{S} poses the issue concerning the presence of extra dimensions. An interesting possibility is the factorisation of the ten-dimensional spacetime into an $AdS_5\times S^5$, a maximally symmetric five-dimensional $AdS$ spacetime factorised from the $S^5$ spatial sphere. Inspection of the physical contents inside the $AdS_5$ geometry has led to the braneworld scenario with a single extra spatial dimension, initially formulated as a thin brane \cite{Ra} and then as thick brane in the presence of scalar fields, as in \cite{gw,De,thick1}. In the thick brane scenario, the presence of source scalar fields that support kinklike configurations capable of generating robust braneworld solutions has increased the interest in the study of localized structures in the real line. In this sense, in the present work we consider a model described by two real scalar fields which are coupled in a very specific way, as recently suggested in \cite{multikink}. The model is different from the Bloch brane model investigated before in Ref. \cite{bloch}, where the scalar fields were described under the presence of standard kinematics. In the present context, in the work \cite{multikink} the coupling between the two fields modified the kinematics of one of them, and this allowed the presence of novel kinklike configurations, engendering interesting internal structure. In the present work, we want to understand how the new solutions may contribute to change the braneworld configuration.

Technically, in the thick braneworld scenario one supposes that the Universe evolves within the five-dimensional warped AdS geometry that is controlled by real scalar fields and the warp factor, which depend only on the extra dimension of infinite extent. In this context, basic issues are then raised, in particular, concerning the robustness of gravity against small fluctuations of the metric, the investigation of the Newtonian limit, the possibility to localize fermion, boson and gauge fields, and the study of the cosmic evolution of the Universe in the braneworld context. These and other related issues have been studied in a diversity of contexts, and we remark that some of the above questions will require further investigations related to the novel thick brane scenario to be developed in this work. The aim of the present work, however, is to provide a novel thick brane scenario which is robust against fluctuations of the metric. Other issues will be investigated elsewhere.

Among the several issues raised within the braneworld context, some very interesting ones concern the phenomenological view  of the brane, as a possible way to respond to the hierarchy problem in particle physics, as introduced before in Refs. \cite{newADD1,newADD2,newADD3}, the brane cosmology \cite{newBCprl,newBC1,newBC2} and the domain-wall/brane-cosmology correspondence \cite{DW1,DW2}. Generically speaking, the hierarchy problem inquires why the masses of the elementary particles are so much lighter than the Planck mass, and brane cosmology concerns the evolution of the Universe within the braneworld scenario. We shall further comment on these issues below, in particular on the cosmological behaviour of a brane-universe, i.e., a three-dimensional space where ordinary matter is confined in the higher dimensional $AdS_5$ spacetime.

 To make the investigation pedagogical, we first deal in Section 2 with the braneworld scenario on general grounds, considering the two source scalar fields to be driven by the Lagrange density suggested in \cite{multikink}. We then move on and in Section 3 we investigate four distinct models, in particular two models that support kinklike solutions which unveil the presence of branes with multiple internal structures, in which the warp factor and energy density depend on the extra dimension symmetrically or asymmetrically. In the asymmetric case, the brane may connect asymptotic $AdS_5$ and Minkowski geometries, or two distinct $AdS_5$ geometries. Interestingly, these possibilities only depend on a single real parameter that controls the asymmetry of the model. We end the work in Section 4, where we summarise the main results and add comments capable of fostering new investigation into the subject.\\

{\bf 2. New braneworld scenario.}
Let us now investigate scalar fields in the braneworld scenario with a single extra dimension of infinite extent, described by the line element
\be \label{elem}
ds^{2}_{5}=e^{2A}\eta_{\mu\nu}dx^{\mu}dx^{\nu}-dy^{2},
\ee
where $e^{2A}$ is the warp factor,  $ A=A(y) $ is the warp function, $ y $ is the extra dimension and $\eta_{\mu\nu}=\text{diag}(+,-,-,-)$ describes the four-dimensional Minkowski metric tensor, with $\mu,\nu=0,1,2,3$. The five-dimensional metric tensor is $g_{ab} = \text{diag}(e^{2A},-e^{2A},-e^{2A},-e^{2A},-1)$, with $a,b=0,1,2,3,4$. In the present study, we deal with the action in the form
\be\label{action}
I=\int dx^{4}dy\sqrt{\vert g\vert}\left(-\frac1{4}{R}+{\cal L}\right),
\ee
where we are using natural units and $4\pi G_{5}=1$, for simplicity. The Lagrangian density describes two real scalar fields and is written as
\be \label{lagran}
{\cal L}=\frac{1}{2}f(\chi)\partial_{a}\phi \partial^{a}\phi+\frac{1}{2}\partial_{a}\chi \partial^{a}\chi-V(\phi,\chi).
\ee
Here, $\phi$ and $\chi$ denote the scalar fields and $f(\chi)$ is in principle an arbitrary function. Notice that the kinematical term related to $\phi$ presents a factor controlled by the $\chi$ field. A similar Lagrangian density was recently considered in Ref.~\cite{multikink} in the study of novel kinklike configurations, that arise in $(1,1)$ spacetime dimensions, where the $\chi$ field simulates, through the function $f(\chi)$, geometrical constrictions that modifies the usual kink profile.

The variation of the previous action with respect to the scalar fields and to the metric tensor leads us to the equations
\bes\label{systeq1}\bal
\label{E1}
&\frac{1}{\sqrt{g}}\,\partial_{a}\!\left(\sqrt{g}\,f\partial^{a}\phi \right) +V_\phi=0,\\
\label{E2}
&\frac{1}{\sqrt{g}}\,\partial_{a}\!\left(\sqrt{g}\,\partial^{a}\chi \right) -\frac{1}{2}f_\chi\,\partial_{a}\phi\partial^{a}\phi+V_\chi=0,\\
\label{E3}
& G_{ab}-2\,T_{ab}=0,
\eal\ees 
where $V_\phi=\partial V/\partial\phi$, $V_\chi=\partial V/\partial\chi$ and $f_\chi=\partial f/\partial\chi$. In Eq.~\eqref{E3}, $G_{ab}$ is the Einstein tensor and $T_{ab}$ represents the energy-momentum tensor. We follow the usual route and consider that the scalar fields are all static, depending only on the extra dimension. In this case, Eqs.~\eqref{E1} and \eqref{E2} become
\bes\label{systeq2}\bal
\label{E11}
& f\phi'' + \left(4A'f + f_\chi\chi'\right)\phi' - V_\phi=0,\\
\label{E22}
&\chi''+4A'\chi'-\frac{1}{2}f_\chi\phi'^{2}-V_\chi=0,
\eal\ees
where the prime represents derivative with respect to $y$, i.e., $\phi^\prime=d\phi/dy$ and $\chi^\prime=d\chi/dy$. Also, the non-vanishing components of Einstein's equations \eqref{E3} are
\bes\label{einst}
\bal \label{einst1}
& A''=-\frac{2}{3}\left(f\phi'^{2}+\chi'^{2}\right), \\ \label{einst2}
& A'^{2}=\frac{1}{6}\left(f\phi'^{2}+\chi'^{2}\right)-\frac{1}{3}V,
\eal
\ees
where $A^\prime=dA/dy$ and $A^{\prime\prime}=d^2A/dy^2$.

As it is now standard in the study of braneworlds, here it is of great interest to find first order equations that solve the above  Eqs.~\eqref{systeq2} and \eqref{einst}. We follow the lines of Refs.~\cite{De,fobrane} and implement a first order formalism, which arise for potentials with the form
\be\label{pot}
V(\phi,\chi)=\frac{1}{2}\frac{W_{\phi}^{2}}{f(\chi)}+\frac{1}{2}W_{\chi}^{2}-\frac{4}{3}W^{2},
\ee
where $W=W(\phi,\chi)$ is in principle an arbitrary function of scalar fields. In this case, we obtain the following first order differential equations
\be \label{first1}
\phi'=\frac{W_{\phi}}{f(\chi)},\qquad\chi'=W_{\chi},
\ee
and
\be \label{first2}
A'=-\frac{2}{3}W\,,
\ee
which are compatible with the scalar field equations of motion \eqref{systeq2} and the Einstein's equations \eqref{einst}. Notice that, in principle, since $W$ presents terms that mix $\phi$ and $\chi$, the equations \eqref{first1} must be solved together as a system. However, a special case arises when $W(\phi,\chi) = g(\phi) + h(\chi)$, such that the first order equations in \eqref{first1} become $\phi'=g_{\phi}/f(\chi)$ and $\chi'=h_{\chi}$. In this situation, the latter equation can be solved independently. It was shown in Ref.~\cite{multikink}, in models with two scalar fields in $(1,1)$ dimensions, that a similar mechanism may be able to simulate geometrical constrictions on kinks, which may give rise to kinks with multiple internal structures, and here we want to investigate this possibility in the braneworld context, in the case of warped five-dimensional geometries with a single extra dimension of infinite extent.

In this model, the energy density can be calculated standardly; it is given by
\be \label{dens}
\rho(y)=e^{2A}\left(\frac{1}{2}f(\chi)\phi'^{2}+\frac{1}{2}\chi'^{2}+V(\phi,\chi)\right).
\ee
With the use of the first order equations, it can be rewritten in the simple form
\be \label{densid}
\rho(y)=(e^{2A}W)^\prime.
\ee
The energy of this system is found by integration of the above expression with respect to the extra dimension, so we get that the energy of the brane associated with the scalar fields is zero, for models described by $W(\phi,\chi)$ that behave adequately asymptotically, since the warp factor $\exp(2A)$ has to vanish asymptotically, to ensure finiteness of the energy of the braneworld configuration.


\subsection*{Stability}

Before going on to investigate specific models, the us now focus on the stability of the brane associated to the model described by Eqs.~\eqref{action} and \eqref{lagran}. We deal with the stability of the gravity sector, considering small perturbations in the metric associated to the line element in Eq.~\eqref{elem}, in the form
\be \label{pertub}
ds^{2}=e^{2A}(\eta_{\mu\nu}+\epsilon h_{\mu\nu})dx^{\mu}dx^{\nu}-dy^{2},
\ee
with $\epsilon$ as a very small parameter and $h_{\mu\nu}=h_{\mu\nu}(x^{\mu},y)$. We use transverse traceless gauge, $h_{\mu\nu}\rightarrow \bar{h}_{\mu\nu}$, to show that the metric fluctuation decouples from the fields~\cite{De}, leading to
\be
(\partial_{y}^{2}+4A'\partial_{y}-e^{-2A}\square)\bar{h}_{\mu\nu}=0,
\ee
where $\square=\eta^{\mu\nu}\partial_{\mu}\partial_{\nu}$ and $\partial_y=\partial/\partial y$. Introducing the $z$ coordinate to make the metric become conformally flat, with the choice $dz=e^{-A(y)}dy$, and defining
$\bar{h}_{\mu\nu}=e^{ip\cdot x}e^{-3A(z)/2}H_{\mu\nu}(z)$, the above equation becomes
\be \label{scho}
\left(-\dfrac{d^{2}}{dz^{2}}+U(z)\right)H_{\mu\nu}(z)=p^{2}H_{\mu\nu}(z),
\ee
where $U(z)$ represents the stability potential, given by
\be \label{U}
U(z)=\frac{9}{4}A_{z}^{2}+\frac{3}{2}A_{zz}\,.
\ee
Here $A_{z}$ and $A_{zz}$ correspond to the first and second derivative of the warp function with respect to the $z$ variable. We remark that the stability equation \eqref{scho} is a Schr\"odinger-like equation, and the stability is ensured when Eq.~\eqref{scho} only admits non negative eigenvalues, i.e., $p^2\geq0$. One can show that Eq.~\eqref{scho} can be written in the form
\be\label{fatoriz}
S^\dagger S H_{\mu\nu}(z) = p^2 H_{\mu\nu}(z),
\ee
with the operators $S$ and $S^\dagger$ given by
\be 
S=\dfrac{d}{dz}-\frac{3}{2}A_{z}\quad\text{and}\quad S^\dagger=-\dfrac{d}{dz}-\frac{3}{2}A_{z}.
\ee
The above factorization shows that the stability equation cannot support states with negative eigenvalues. Hence, the brane is stable under small fluctuations of the metric.\\

\begin{figure}[t]
\centering{
{\includegraphics[width=6.8cm]{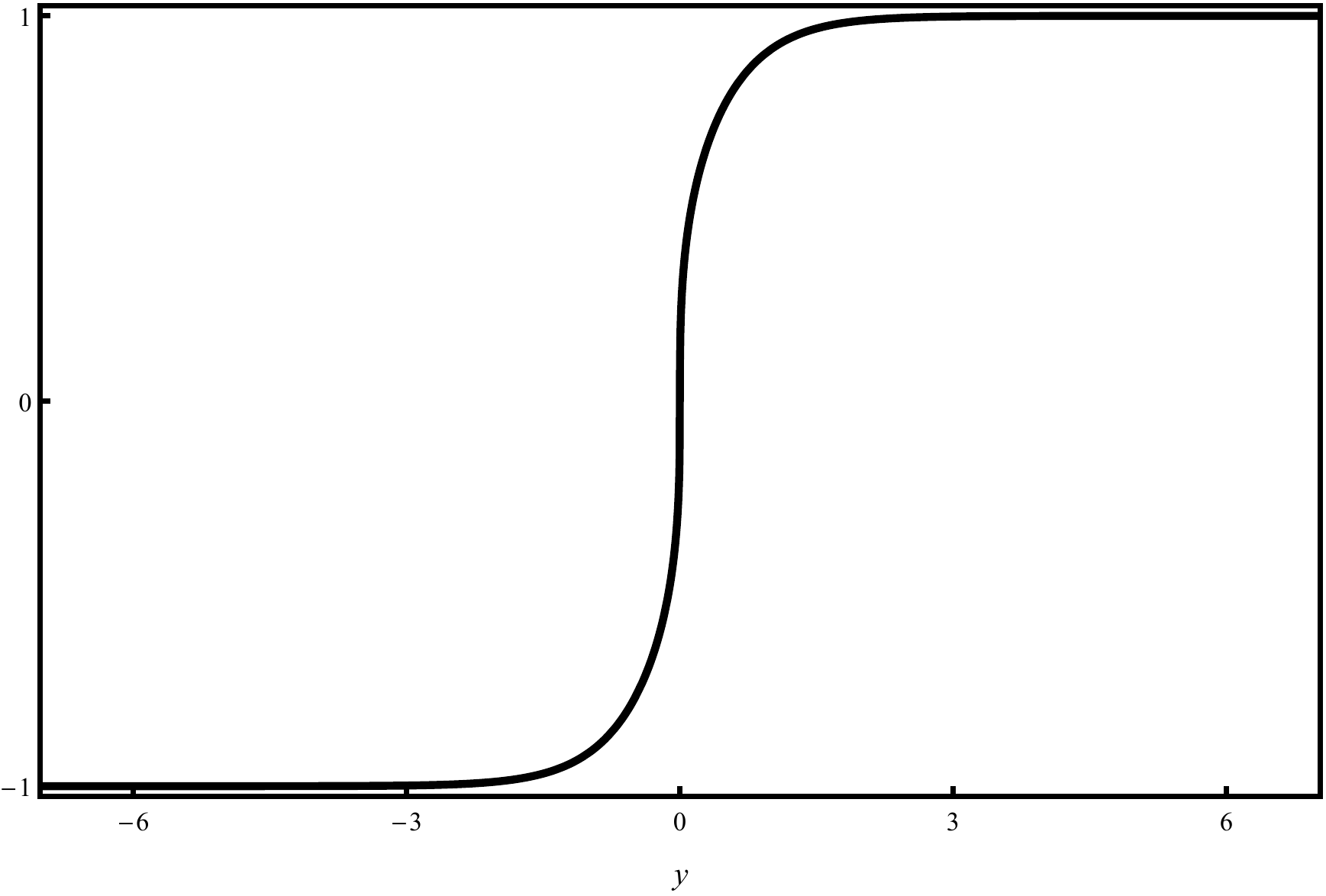}}
{\includegraphics[width=6.8cm]{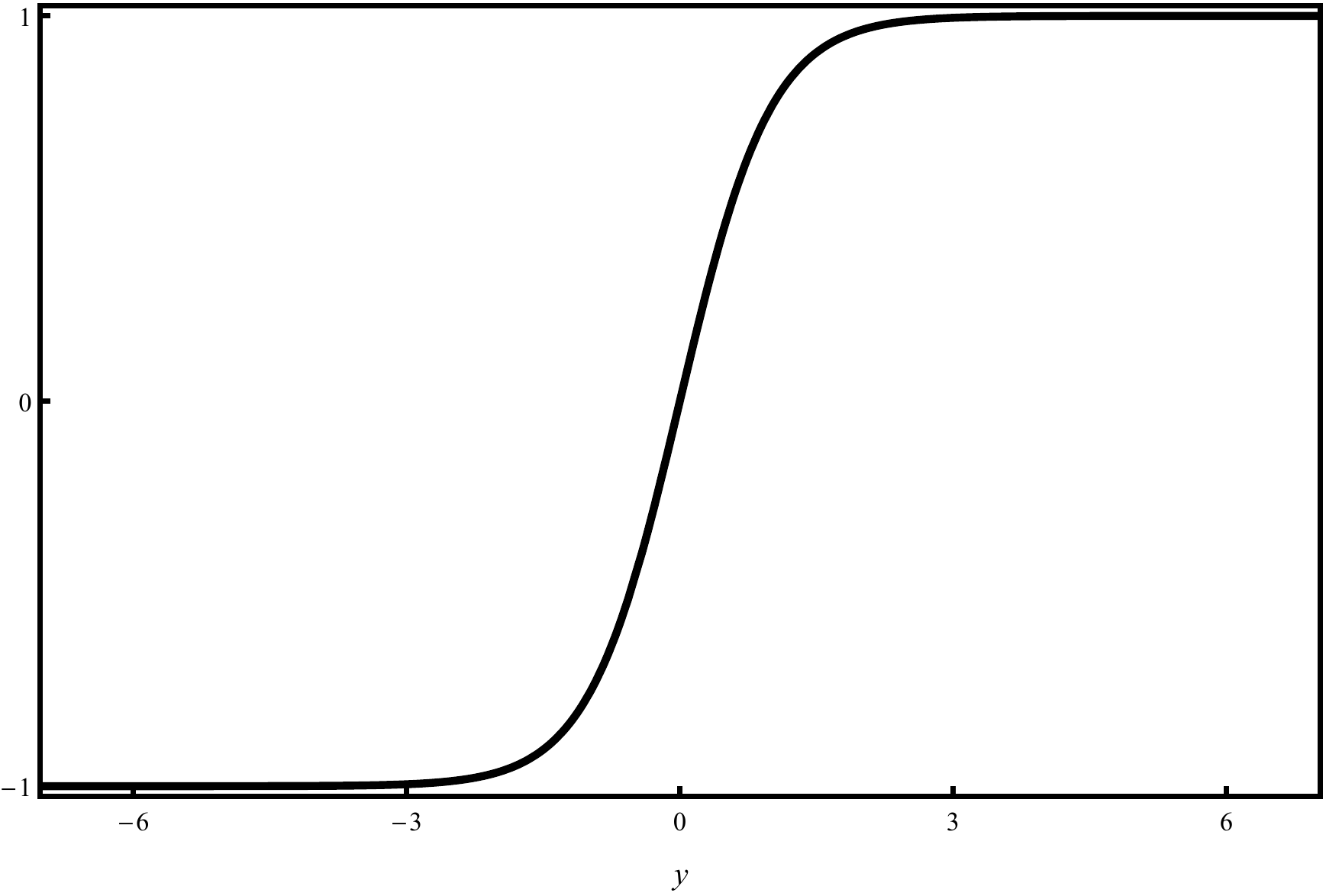}}
{\includegraphics[width=6.8cm]{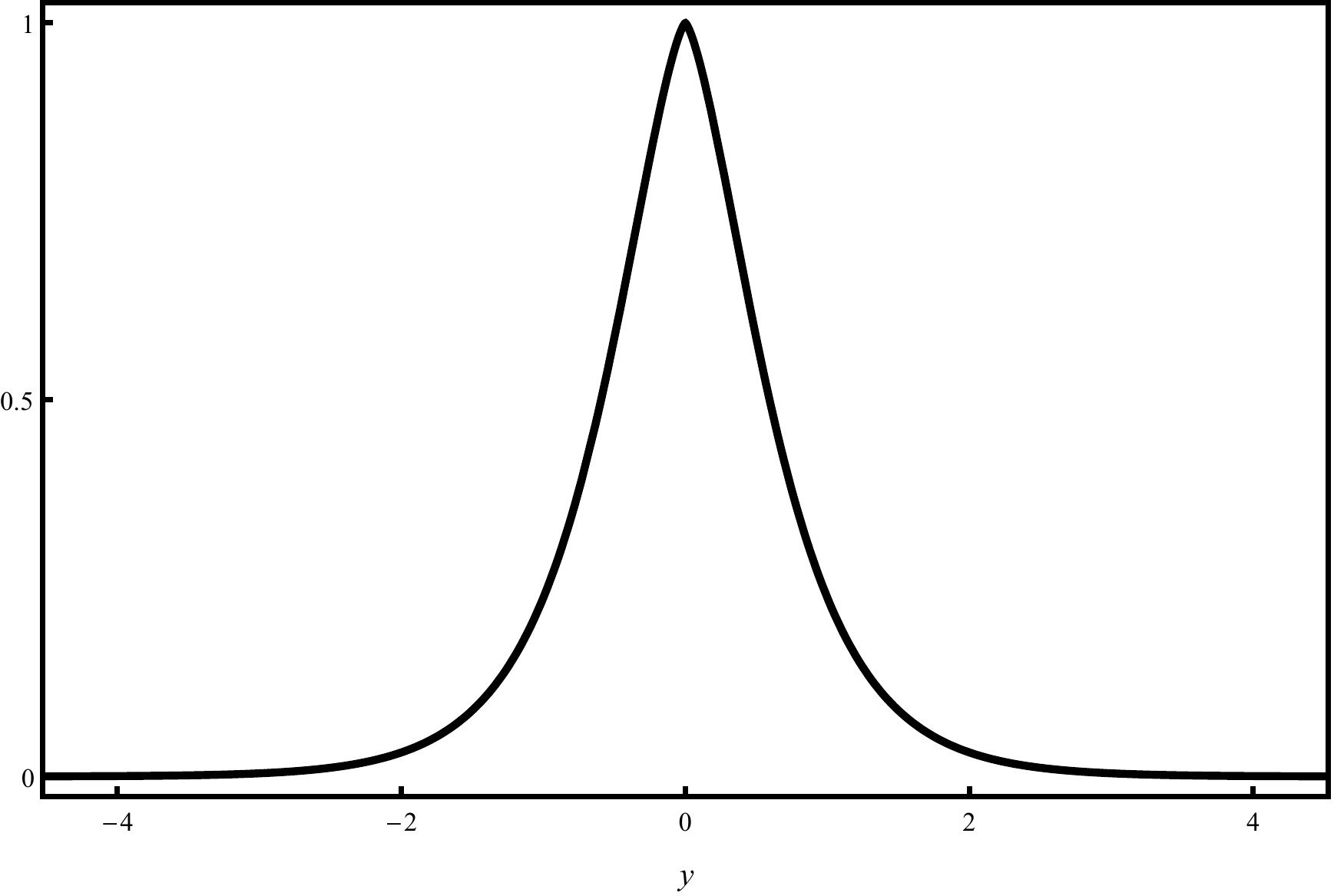}}
{\includegraphics[width=6.8cm]{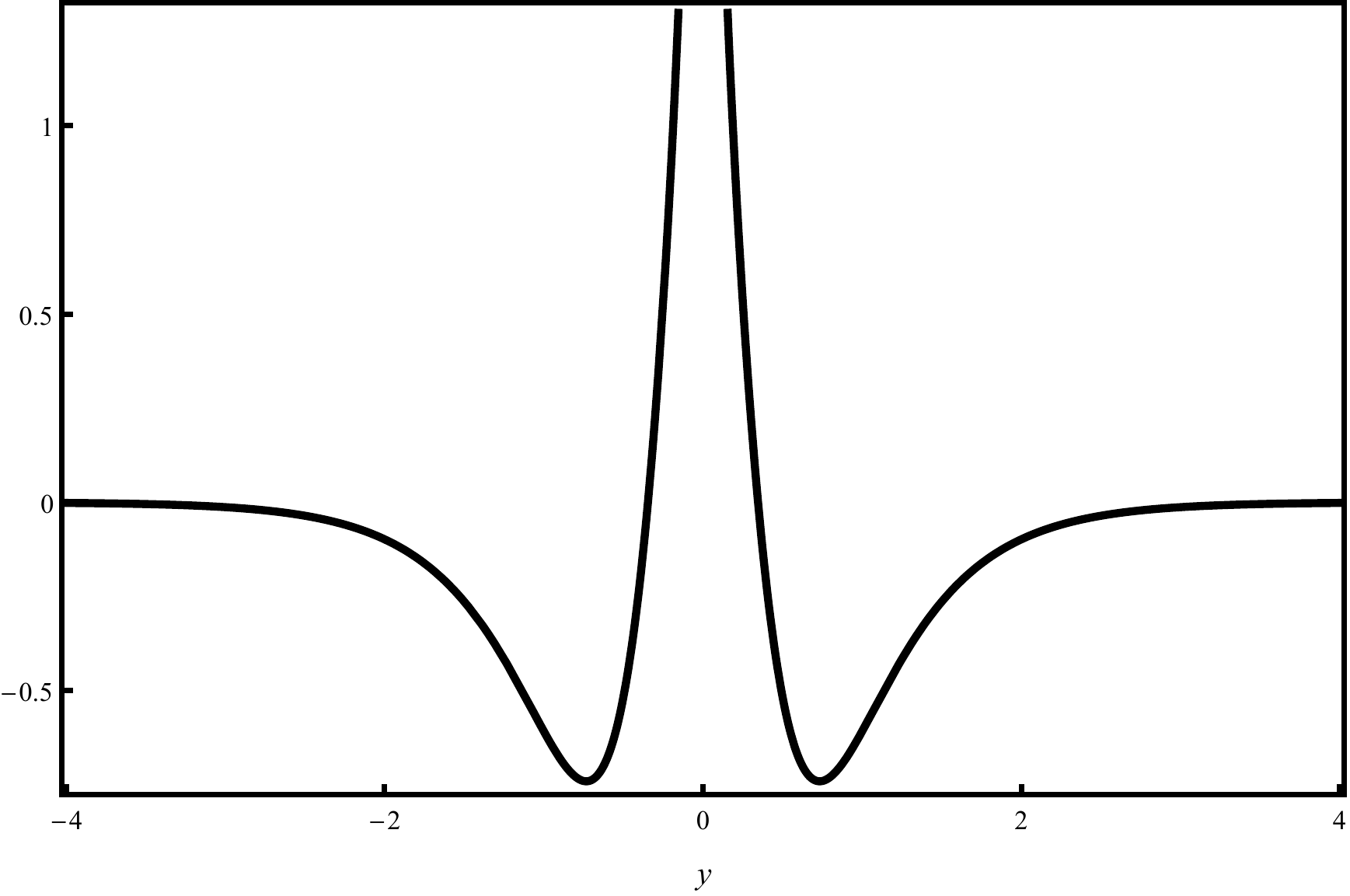}}}
\caption{From top to bottom, we depict the solutions $\phi(y)$ and $\chi(y)$ for $\alpha=1$, the warp factor and the energy density associated to model 1.}\label{Fig1}
\end{figure}

{\bf 3. Specific Models.}\label{sec3}
To see how the general procedure works for building thick brane solutions, let us now illustrate it with several models, which we describe below. Since we are considering $W(\phi,\chi)=g(\phi)+h(\chi)$, in this case the $\chi$ field can be investigated independently and we then first choose
\be\label{h}
h(\chi)=\alpha\left(\chi-\frac{\chi^3}{3}\right),
\ee  
where $\alpha$ is a real parameter. This choice shows that the $\chi$ field works as a prototype of the Higgs field: it engenders a double well potential of the $\chi^4$ type, and it has the interesting standard behavior which was carefully investigated before in Ref. \cite{De} within the thick brane scenario. The corresponding first order equation is solved by
\be\label{chi}
\chi(y)=\tanh(\alpha y),
\ee
which is well-understood in the thick brane context \cite{De}. The main motivation here is to use a well-known model for the field $\chi$, to see how it may modify the behavior of the other field, $\phi$, in the context of the model under investigation in the present work. 

We further notice that the first order equation for $\chi$ in Eq. \eqref{first1} also admits the uniform solutions $\bar{\chi}_\pm=\pm1$; thus, when $\chi$ assumes one of these two values, the contribution of $\chi$ for the model \eqref{lagran} disappears, if one further imposes that $f(\chi=\pm1)$ is a positive real constant, which can be absorbed by a scaling in the scalar field. We then see that it is of interest here to choose the function $g(\phi)$ to still provide a robust model of thick brane with a single real scalar field. The above reasonings set the general guidance for the construction of models, and we now move on and turn attention to the specific models which we introduce and investigate below.
 
\subsection*{Model 1}
The first model is described by the function
\be \label{super1}
g(\phi)=\left(\phi-\frac{\phi^{7}}{7}\right)\,,
\ee
and the choice
\be
f(\chi)=3\sqrt[3]{\chi^{2}}.
\ee
We see that $f(\pm1)$=3, so it has the desired property, and the derivative of $g$ gives $1-\phi^6$, which also leads to a double well potential with minima at $\bar{\phi}_{\pm}=\pm 1$, and so capable of supporting a thick brane configuration. In the case with $\chi=\tanh(\alpha y)$, the first order equation for the scalar field $\phi$ which is in \eqref{first1} becomes
\be \label{dif1a}
\phi^\prime=\frac{(1-\phi^{6})}{3\sqrt[3]{\tanh^2(\alpha y)}}.
\ee
We take $\alpha=1$ and get the solution
\be \label{sol1}
\phi(y)=\sqrt[3]{\tanh(y)},
\ee
which can be seen in the top panel in Fig.~\ref{Fig1}. We notice that the $\phi$ solution presents infinite derivative at the origin, which is a consequence of the $f(\chi)$ function fed by the $\chi$ solution. Thus, our method allows us to modify the derivative of the $\phi$ field that plays an important role in the warp factor and energy density of the brane, as we show below.

In order to calculate the warp function, we use Eq.~\eqref{first2}, which reads
\be\label{dif1b}
\begin{aligned}
	A^\prime &=-\frac{2}{3}\bigg(\tanh^{\frac13}(y)+\tanh(y)-\frac{1}{7}\tanh^{\frac73}(y)\\ 
&\hspace{3.9mm}-\frac{1}{3}\tanh^{3}(y)\bigg).
\end{aligned}
\ee
From the above equation, we can see that $A^\prime(0)=0$ and $A''(0)\to-\infty$. This makes the warp function to develop a sharp peak at the origin, a phenomenon which is similar to the behavior of the thin brane scenario proposed in Ref.~\cite{Ra}. In this sense, the above model somehow simulates a thin brane, even though we are considering  topological structures that appear from the two field model used to generate the brane,
We were able to solve the above equation and calculate the warp function analytically with the condition $A(0)=0$. However, since the full expression is very long, we omit it here. In the two bottom panels of Fig.~\ref{Fig1} we depict the warp factor and the energy density of the model. Notice that, as we stated before, at $y=0$, the warp factor shows a sharp peak, as in a thin brane, and the smooth behavior of a thick brane outside the origin. This hybrid behavior is the opposite of the one found in Ref.~\cite{hybrid}, where the brane behaves as a thick brane inside a compact space around the origin and as a thin brane outside the compact space. Moreover, the total energy vanishes, as expected.

\subsection*{Model 2}
\begin{figure}[t]
\centering{
{\includegraphics[width=6.8cm]{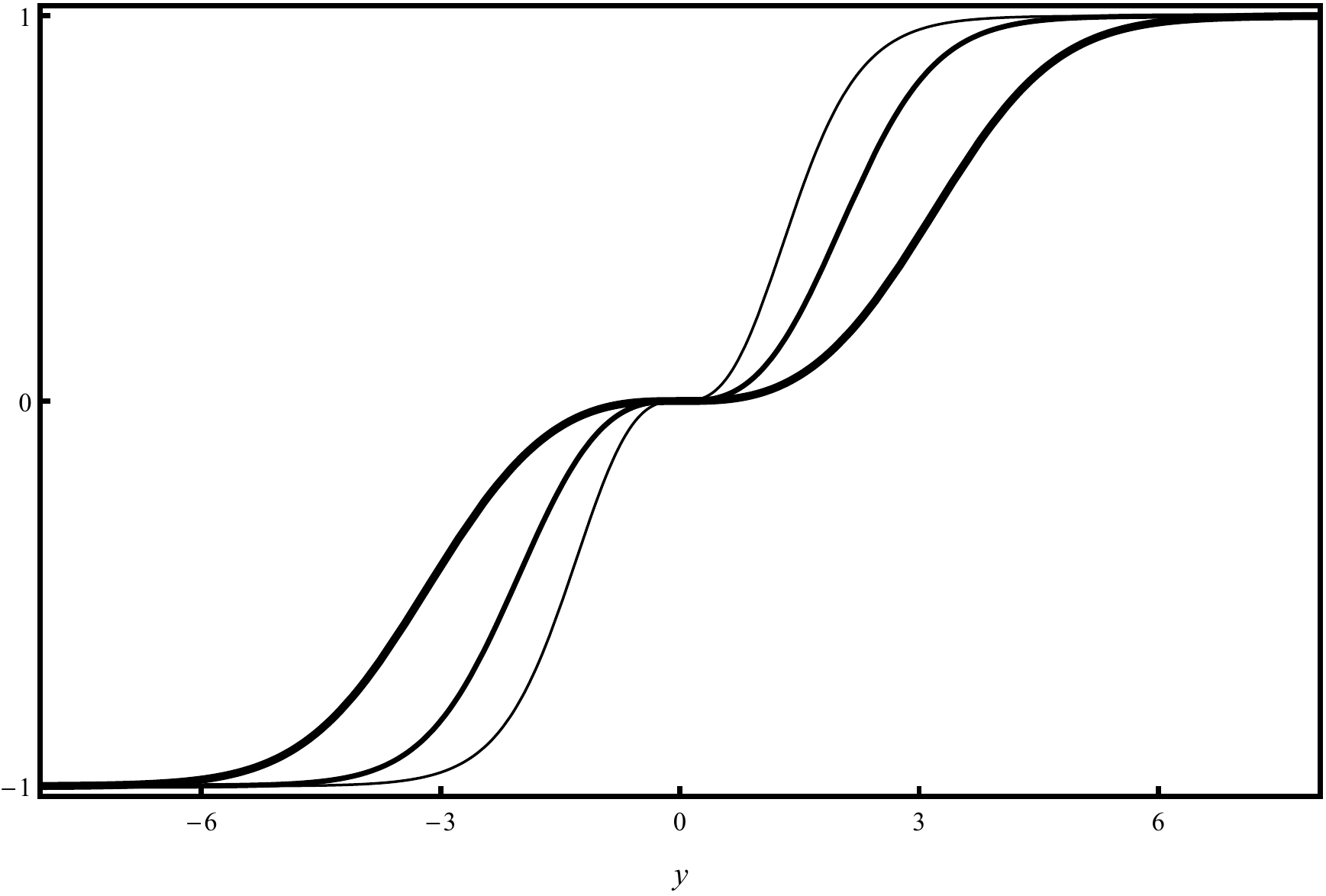}}
{\includegraphics[width=6.8cm]{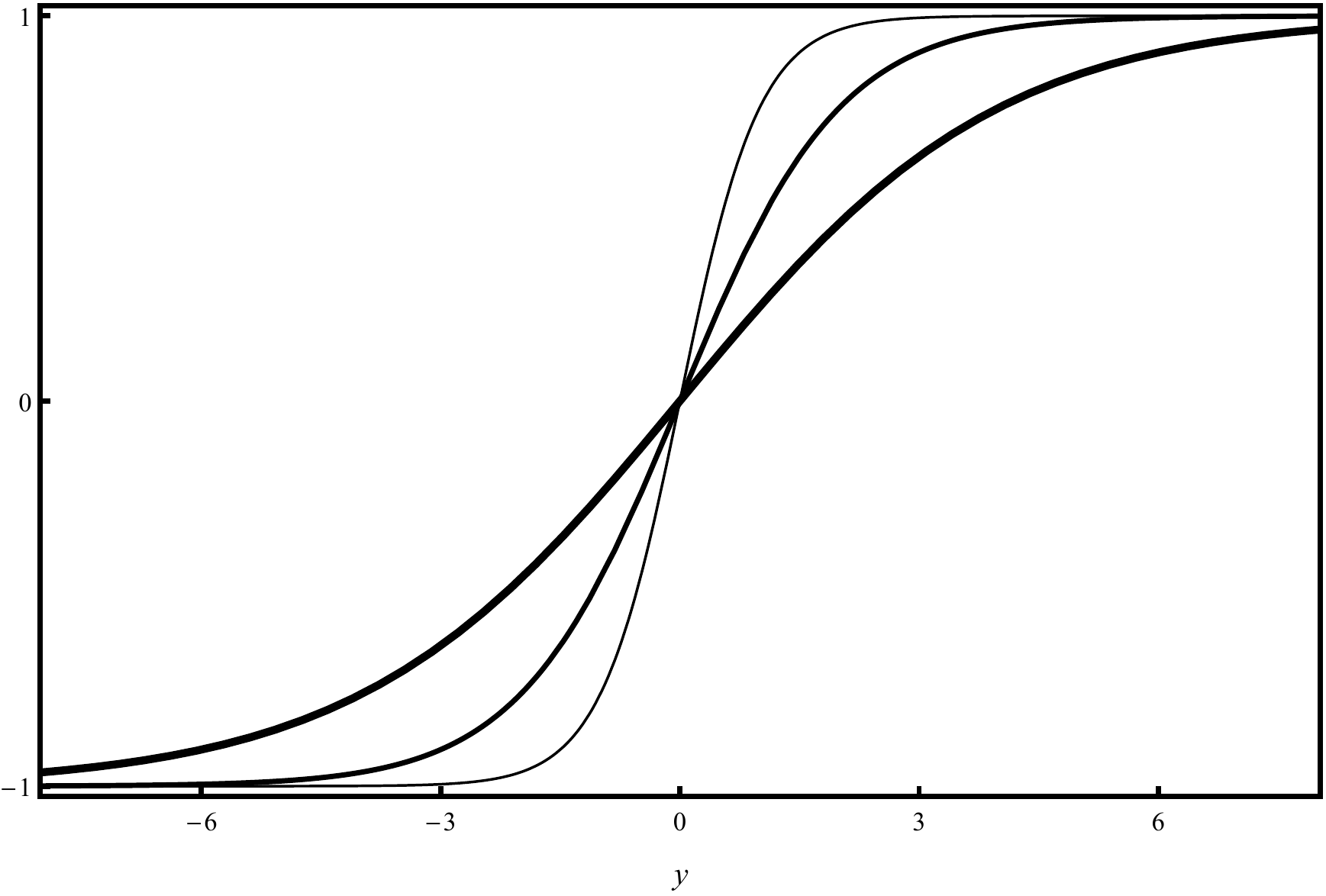}}
{\includegraphics[width=6.8cm]{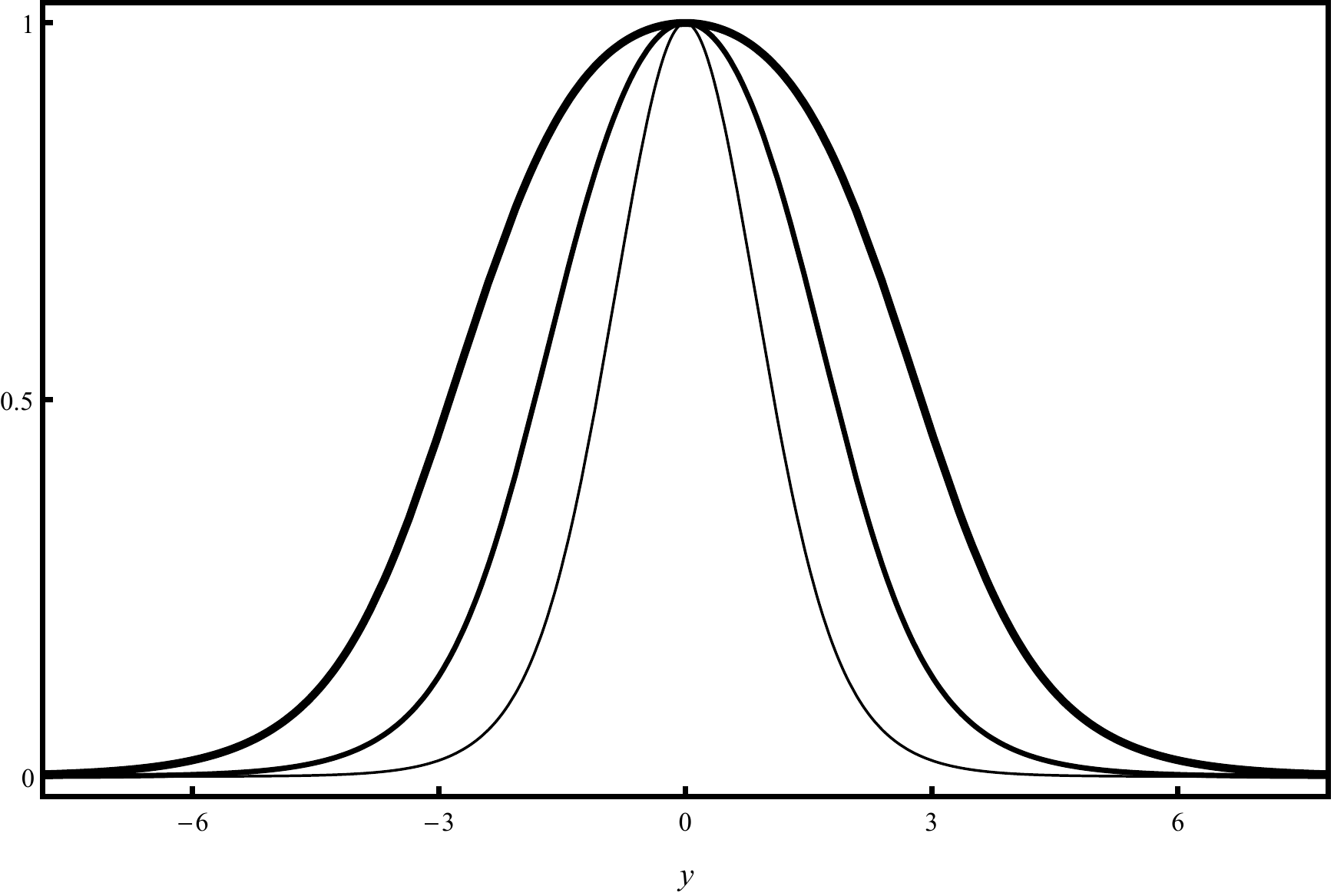}}
{\includegraphics[width=6.8cm]{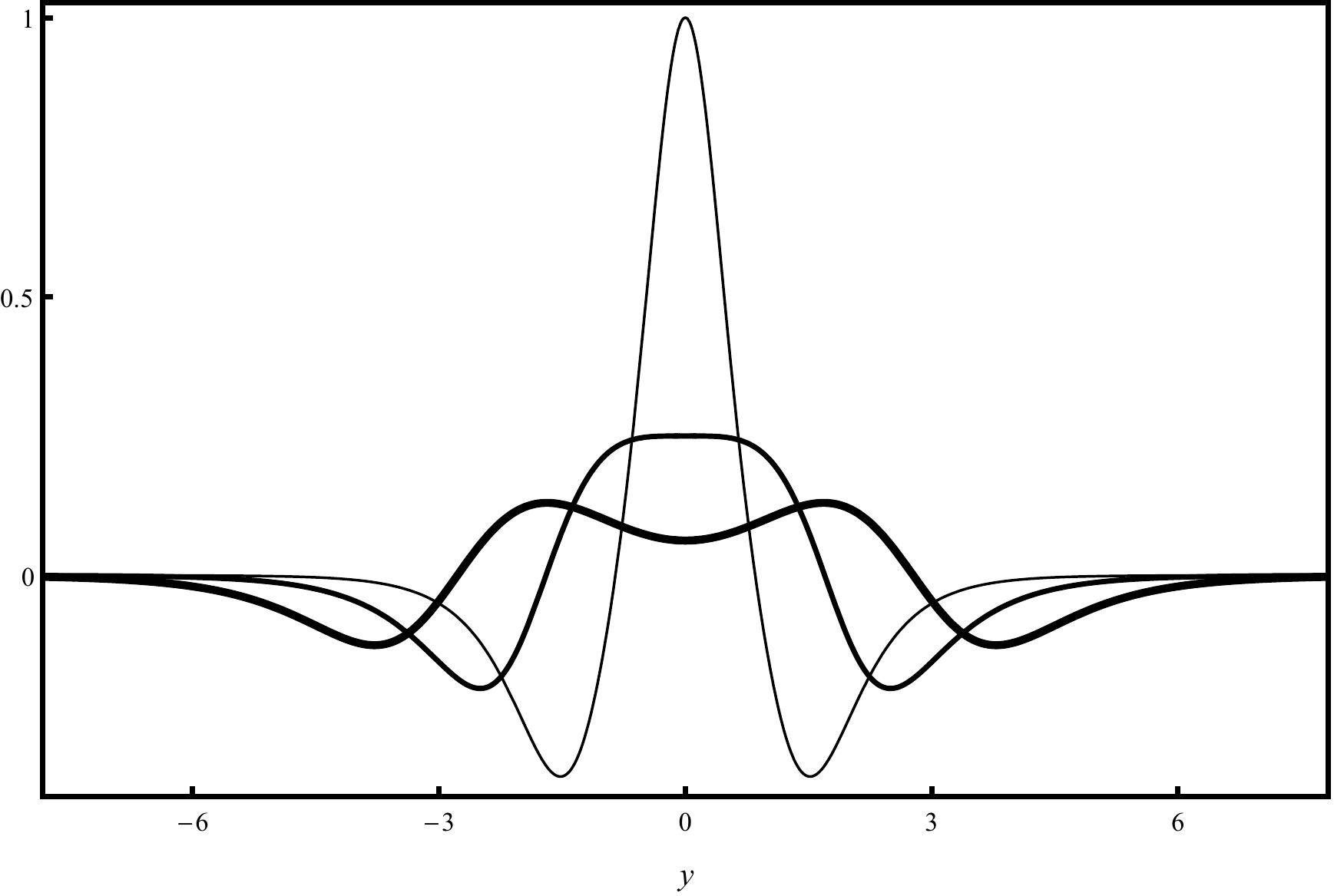}}}
\caption{From top to bottom, we depict the solution $\phi(y)$, the solution $\chi(y)$, the warp factor and the energy density associated to model 2. We take $\alpha=1/4,1/2$ and $1$, with the thickness of the lines increasing as $\alpha$ decreases.}
\label{Fig2}
\end{figure}

The next model we investigate is described by 
\be
\label{super2}
 g(\phi)=\left(\phi-\frac{\phi^{3}}{3}\right),
 \ee
 and
 \be
 f(\chi)=\frac{1}{\chi^{2}}\,,
\ee
We notice that $f(\pm 1)=1$ and that $g(\phi)$ describes the well-known $\phi^4$ model, which is also well-understood in the thick brane context \cite{De}. The motivation here is then to grasp the importance of the function $f(\chi)$ to modify the behavior of the braneworld solution in the thick brane scenario. This model is also motivated by the one found Ref.~\cite{multikink}, which simulates the geometrical constrictions that give rise to double kink profiles in the magnetization of specific magnetic materials \cite{KCM1}. For the function given above, the first order equation for the field $\phi$ which is shown in Eq. \eqref{first1} now becomes
\be \label{dif2a}
\phi^\prime=(1-\phi^{2})\chi^{2}.
\ee
This equation supports the kinklike solution
\be \label{sol2}
\phi(y)=\tanh(\xi(y)),
\ee
with $\xi(y)=y-\tanh(\alpha y)/\alpha$. Its profile is depicted in the top panel of Fig.~\ref{Fig2}, which also shows the profile of the field $\chi(y)=\tanh(\alpha y)$. Notice that, near the origin, the $\phi$ kink shows a plateau whose width is controlled by $\alpha$. This occurs because, for $y\approx0$, we have $\phi\propto \alpha^2 y^3$; thus, as $\alpha$ decreases, the plateaux gets wider.

The warp function is driven by Eq.~\eqref{first2}, which now reads
\be
\begin{aligned}
A^\prime &=-\frac23\bigg(\tanh(\xi(y))-\frac{\tanh^{3}(\xi(y))}{3} \\&\hspace{3.9mm}+\alpha \bigg(\tanh(\alpha y)-\frac{\tanh^{3}(\alpha y)}{3}\bigg)\bigg).
\end{aligned}
\ee
We have not been able to find the warp function analytically for this model. So, we perform the calculations numerically for some values of $\alpha$. As before, we take $A(0)=0$. In the two bottom panels of Fig.~\ref{Fig2} we display the warp factor and the energy density of the model. Notice that the warp factor gets wider as $\alpha$ decreases. In the energy density, we see that $\alpha$ gives rise to an internal structure, which appears for $\alpha\leq\alpha_{crit}\approx1/2$. We recall that the presence of internal structure in branes was studied previously in Ref.~\cite{branestructure}, but here we included a parameter that can be used to control the internal structure of the brane. As we can see, as $\alpha$ gets smaller, the internal structure gets larger. This is an issue of current interest, since the internal structure modifies both the warp factor and the energy density of the brane, so it may change conformational aspects of the brane that can, for instance, modify the Newtonian limit, the entrapment of boson, fermion, and gauge fields \cite{fl1,fl2,gl0,gl1}, and the associated cosmic evolution and/or the domain-wall/brane-cosmology correspondence \cite{newBC1,newBC2,DW1,DW2}.

\begin{figure}[t]
\centering{
{\includegraphics[width=6.8cm]{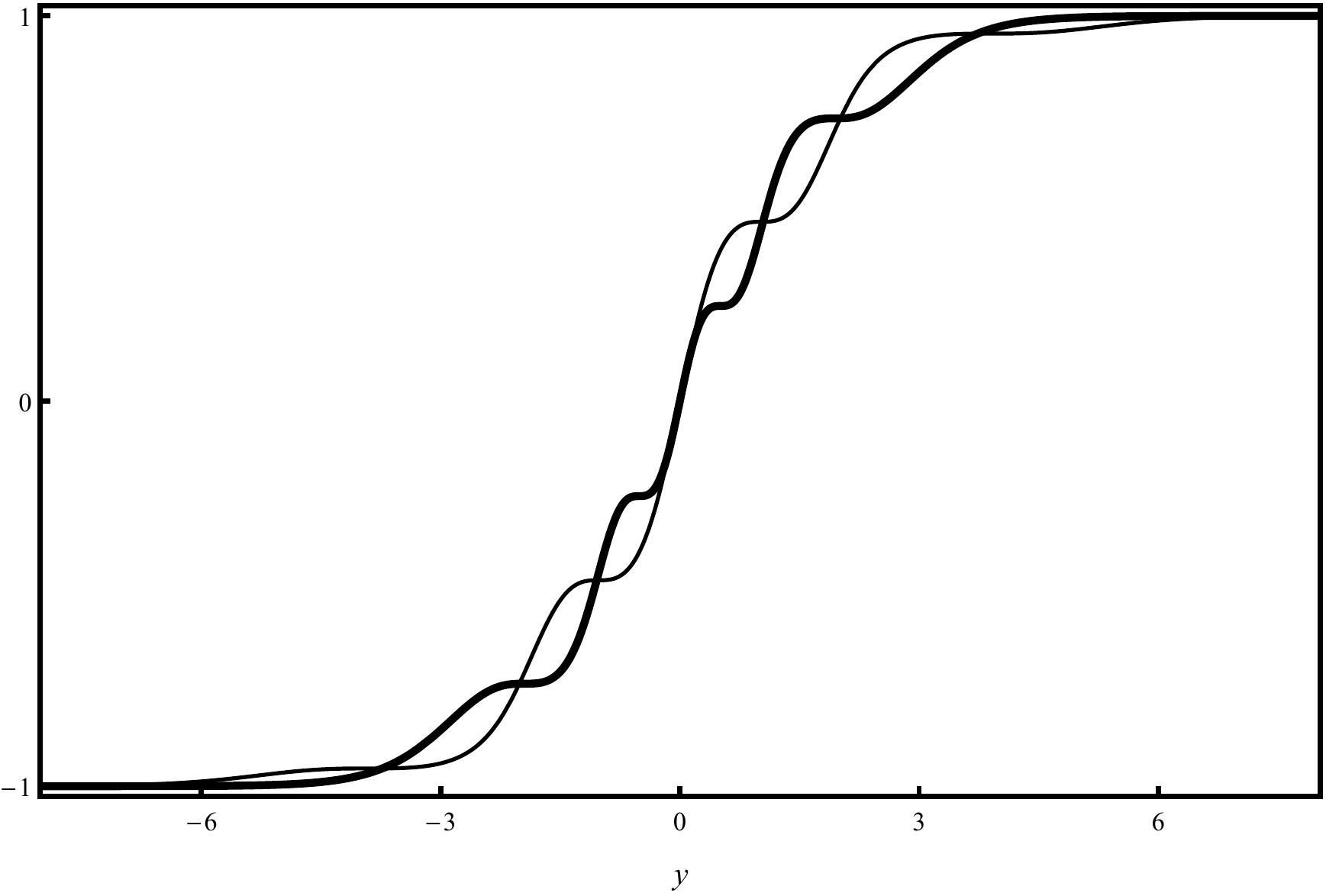}}
{\includegraphics[width=6.8cm]{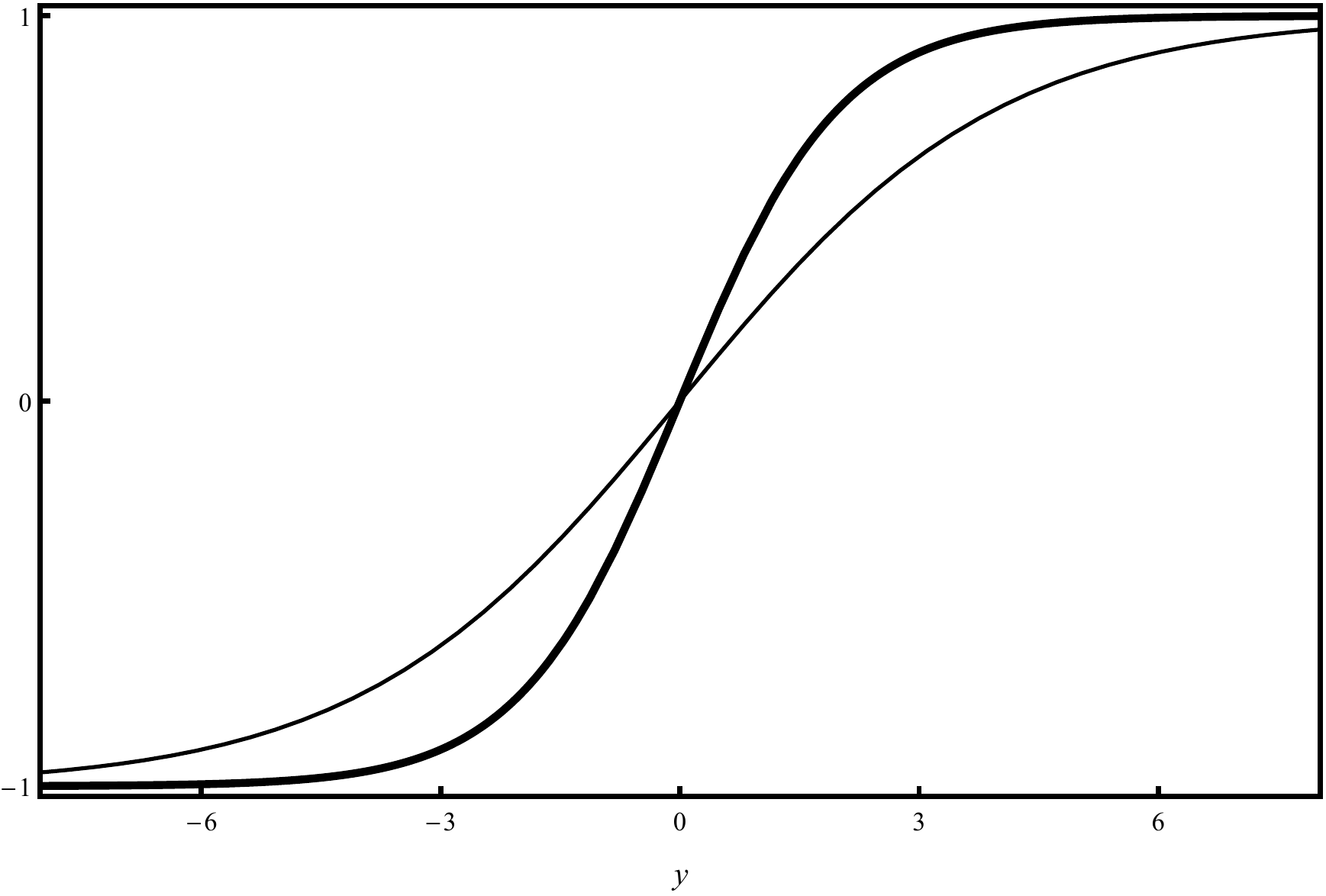}}
{\includegraphics[width=6.8cm]{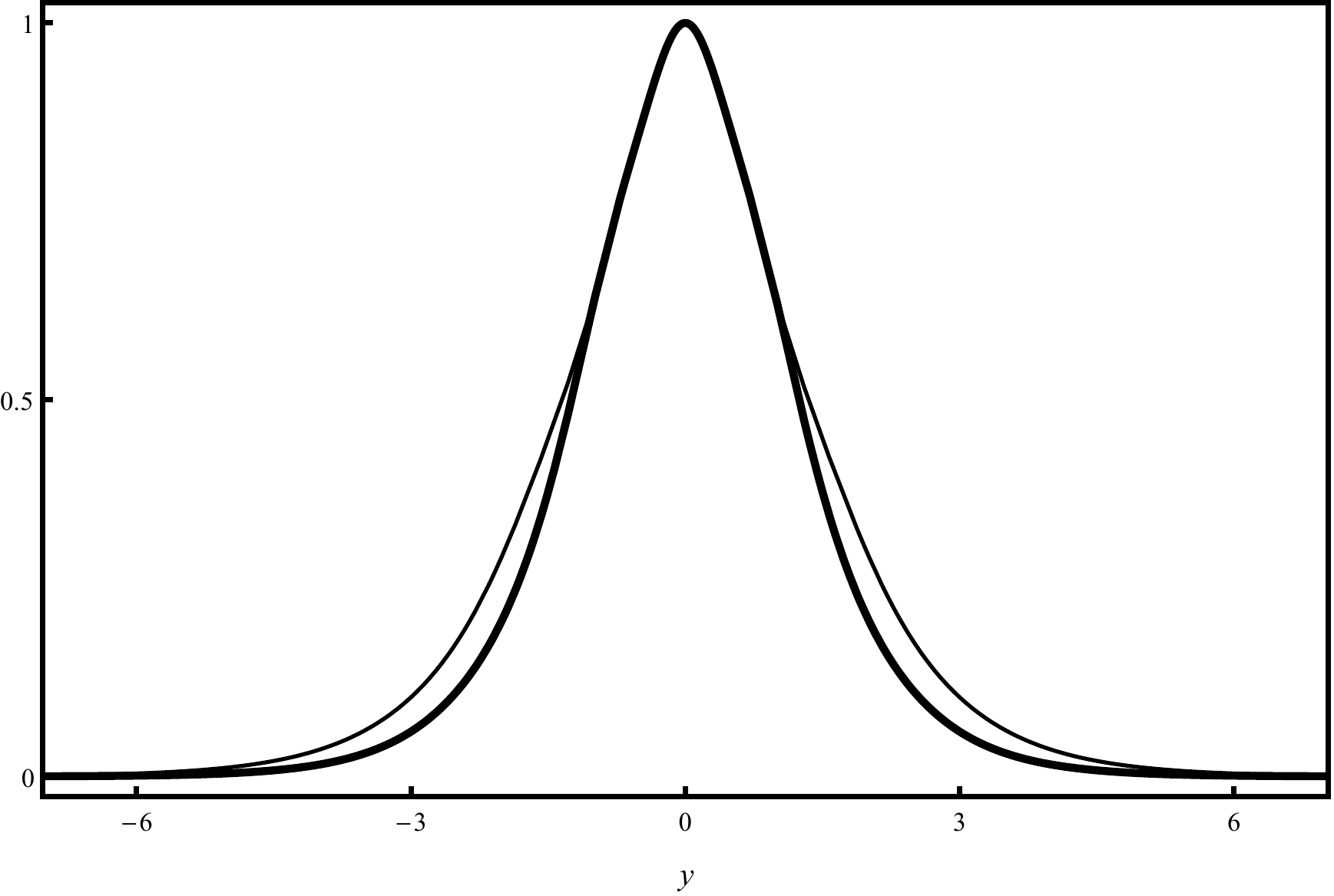}}
{\includegraphics[width=6.8cm]{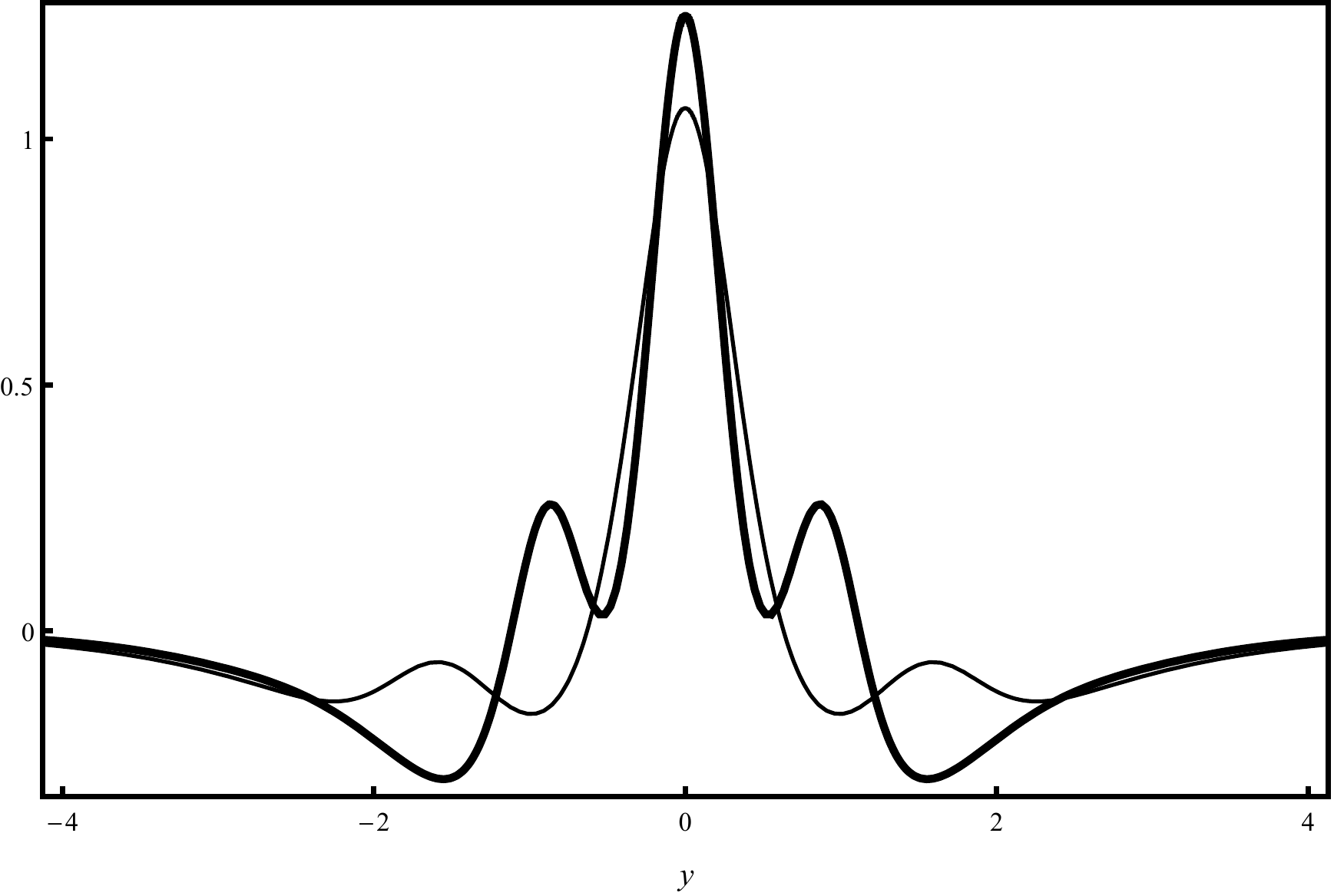}}}
\caption{From top to bottom, we depict with $n=2$ and $\alpha=1/2$ (thick line) and $1/4$ (thin line), the solution $\phi(y)$, the solution $\chi(y)$, the warp factor and the energy density associated to model 3.}
\label{Fig3}
\end{figure}

\subsection*{Model 3}
The third model is described by the same $g(\phi)$ and $h(\chi)$ of the previous model, but here we choose another $f(\chi)$, of the form
\be \label{super3}
f_n(\chi)=\frac{1}{\cos^{2}(n\pi \chi)},
\ee
with $n$ integer. We notice that the above function is periodic; also, the cosine vanishes when its argument is equal to $k\pi/2$, with $k$ being an odd integer. As shown in Ref.~\cite{multikink}, this feature creates several plateaux in the kinklike solution, controlled by $n$. Here, we want to investigate its effect on the brane. Before doing that, however, we notice the uniform solutions $\bar \chi_{\pm}=\pm1$ lead to $f_n(\bar\chi_\pm)=1$, so the model reduces to the well-known thick brane model studied before in \cite{De}, controlled by the Higgs-like source field with the $\phi^4$ potential. Moreover, when the $\chi$ field gets the kinklike form $\chi(y)= \tanh(\alpha y)$, it may induce interesting modification on the $\phi$ field, and also on the warp factor and energy density of the brane. To see how this works, we consider the first order equation for $\phi$, which now has the form
\be\label{dif3a}
\phi^\prime=(1-\phi^{2})\cos^{2}(n\,\pi\, \tanh^2(\alpha y)),
\ee
which has the solution
\bes\label{sol3}\bal
\label{sol3a}
\phi(y)&=\tanh\,(\eta(y)),\\
\label{sol3b}
\eta(y)&=\frac{y}{2}+\frac{1}{4\alpha}\left(\Ci(\xi^{+}_{n}(y))-\Ci(\xi^{-}_{n}(y))\right),
\eal\ees 
where $\Ci(x)$ is the cosine integral function and $\xi^{\pm}_{n}(y)=2n\pi(1\pm \tanh(\alpha y))$. The above solution is depicted for $n=2$ and $\alpha=1/2$ and $1/4$ in the top panel of Fig.~\ref{Fig3}, together with the $\chi(y)$ which appears just below. Notice that we now have a kinklike structure with $2n$ plateaux.

To calculate the warp function, one must use Eq.~\eqref{first2}, which has the form
\be
\begin{aligned}
	A^\prime &=-\frac23\bigg(\tanh(\eta(y))-\frac{\tanh^{3}(\eta(y))}{3} \\
	         &\hspace{3.9mm}+\alpha\bigg(\tanh(\alpha y)-\frac{\tanh^{3}(\alpha y)}{3}\bigg)\bigg).
\end{aligned}
\ee
Unfortunately, we have not been able to find the analytical solution for $A(y)$ in this model. We then perform the calculations numerically with the condition $A(0)=0$. In the two bottom panels of Fig. \ref{Fig3} we depict the warp factor and the energy density, respectively, for $n=2$ and $\alpha=1/2$ and $1/4$. Notice that, even though the warp function is not significantly modified by the plateaux in the solutions, one can see the energy density now gets interesting internal structures. Furthermore, there is an odd number of maxima in the energy density for finite $y$; they are controlled by $n$, counted as $2n-1$.

In this symmetric braneworld model we have two parameters available, $\alpha$, real, and $n$, integer. We believe they can be used in several applications, for instance, in the investigation of the Newtonian limit of gravity and the entrapment of boson, fermion, and gauge fields inside the brane \cite{fl1,fl2,gl0,gl1} and the associated cosmic evolution and/or the domain-wall/brane-cosmology correspondence \cite{newBC1,newBC2,DW1,DW2}.
\subsection*{Model 4}

\begin{figure}[t]
\centering{
{\includegraphics[width=6.8cm]{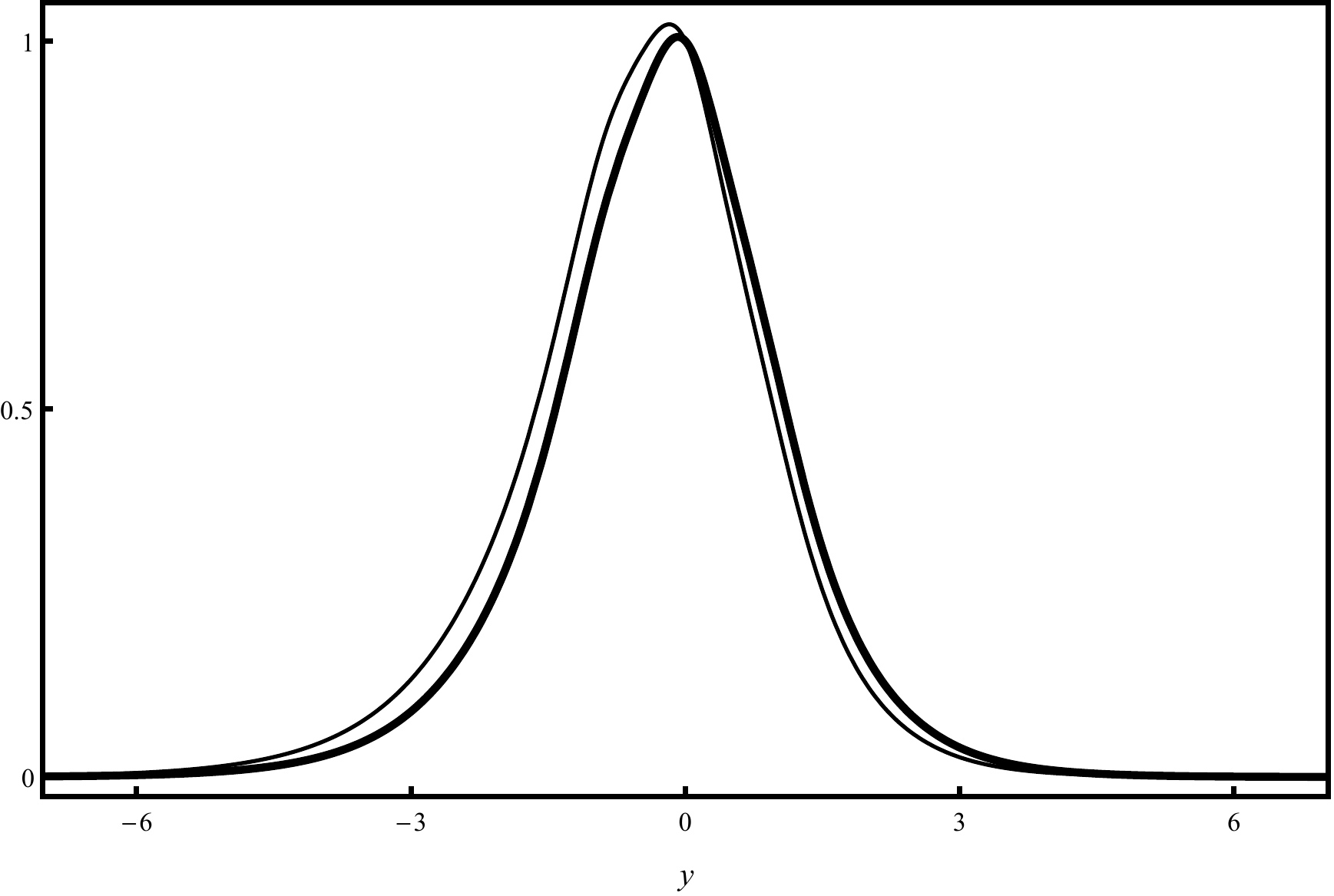}}
{\includegraphics[width=6.8cm]{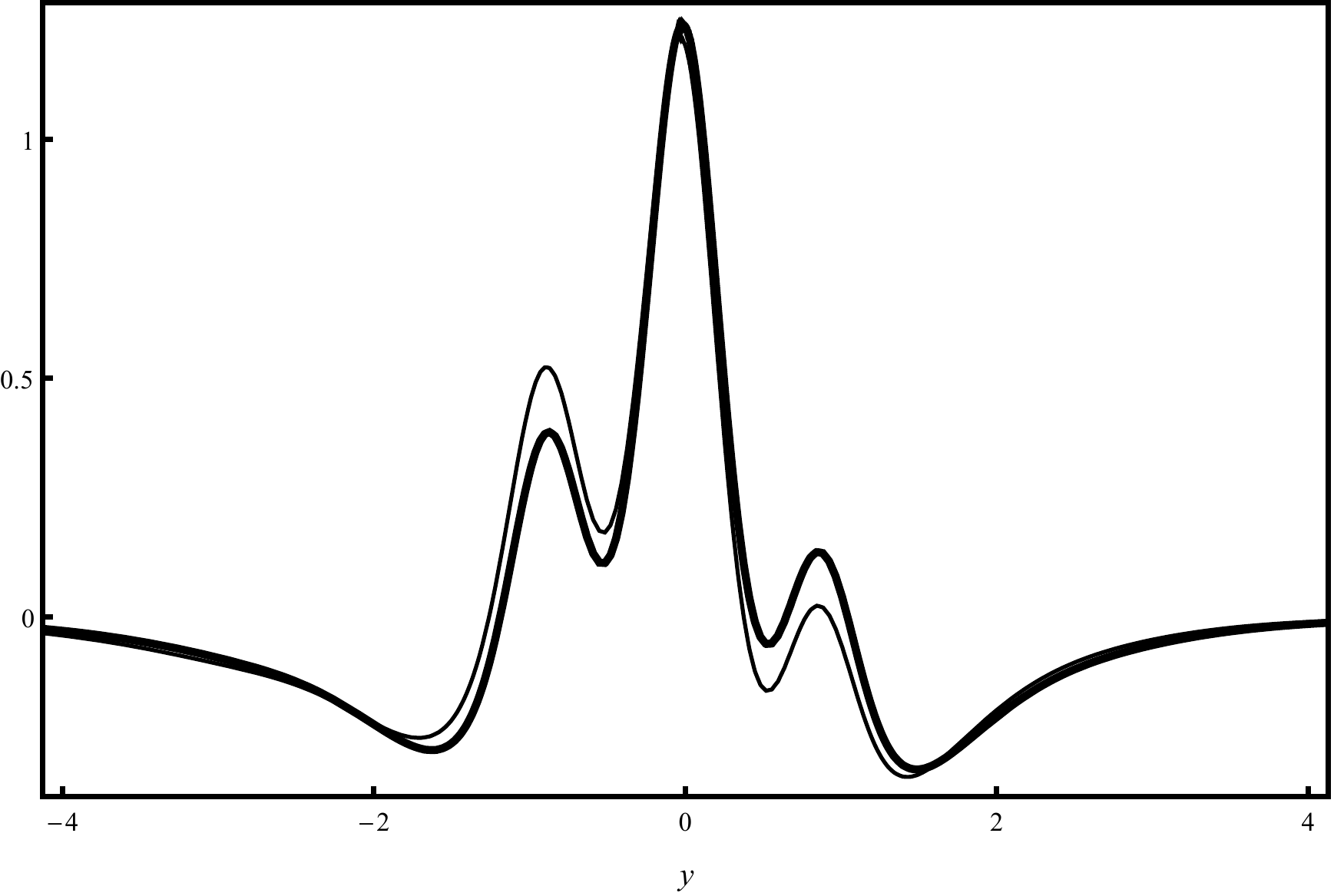}}}
\caption{From top to bottom, we depict the warp factor and the energy density associated to model 4, using $n=2$, $\alpha=1/2$ and $k=0.1$ (thick line) and $0.2$ (thin line).}
\label{Fig4}
\end{figure}

Let us now consider the model with the same $h(\chi)$ defined in Eq. \eqref{h},
the same $g(\phi)$ defined in Eq. \eqref{super2} and the same $f(\chi)$ defined in Eq. \eqref{super3}, but now with a real parameter $k$ added to $W(\phi,\chi)$, changing it to the new form
\be
 W_{\alpha,k}(\phi,\chi)=\left(\phi-\frac{\phi^{3}}{3}\right)+\alpha\left(\chi-\frac{\chi^{3}}{3}\right)+k.
\ee
The addition of $k$ is inspired in the recent work \cite{BF}, which develops the possibility of inducing asymmetric behavior to the brane; see, e.g., Refs. \cite{A1,A2,A3} for other related issues. In the present work, we want to investigate how the parameter $k$ works in the new model. We first notice that the constant $k$ does not change the equations of motion for both $\phi$ and $\chi$, so they behave as in the previous model. However, the warp function $A=A(y)$ feels the presence of $k$ through the equation 
\be
\begin{aligned}
	A^\prime &=-\frac23\bigg(\tanh(\eta(y))-\frac{\tanh^{3}(\eta(y))}{3} \\
	         &\hspace{3.9mm}+\alpha\bigg(\tanh(\alpha y)-\frac{\tanh^{3}(\alpha y)}{3}\bigg)+k\bigg).
\end{aligned}
\ee
We solve this equation numerically: for $n=2$ and $\alpha=1/2$, and for $k=0.1$ and $0.2$ the results are shown in Fig. \ref{Fig4}. They clearly show that $k$ induces an asymmetric behavior to the brane, with the asymmetry being controlled by $k$. The bigger the $k$, the larger the asymmetry. 

Another issue related to the addition of $k$ concerns the presence of asymmetric five-dimensional cosmological constants, as explored before in \cite{BF}. In the present case, asymptotically we can define 
\be
\Lambda_{5\pm}=-\frac43( k+W_\alpha(\phi_\pm,\chi_\pm))^2
\ee
where $+$ and $-$ are used to identify the asymptotic behavior as the extra dimension $y$ tends to $\pm\infty$. We consider the case $\alpha=1/2$ to write
\be
\Lambda_{5\pm}=-\frac43(k\pm1)^2,
\ee
with $k$ constrained to vary in the interval [-1,1] to avoid breaking the brane model robustness \cite{BF}. In the case of $k=\pm1$, the brane lives inside asymptotically $AdS_5$ and Minkowski geometries. Also, for $|k|\in(0,1)$, the brane connects two distinct $AdS_5$ geometries. 

In this asymmetric model we have three parameters: $\alpha$, real, $n$, integer, and $k$ real. We believe they may be used in applications such as the Newtonian limit, the entrapment of boson, fermion, and gauge fields \cite{fl1,fl2,gl0,gl1}, the associated cosmic evolution, and the domain-wall/brane-cosmology correspondence \cite{newBC1,newBC2,DW1,DW2}.\\

{\bf 4. Conclusion.} In this work we have studied the modelling of branes in warped five-dimensional geometries by two scalar fields coupled as in the Lagrange density in Eq.~\eqref{lagran}. We have calculated the equations of motion that describe our system and, also, we have shown that the energy of the brane is null. Since the equations of motion are of second order, to simplify the problem, we have developed a first order formalism. In this framework, the potential is required to have a specific form, in which the coupling between the scalar fields appear. The first order formalism shows that the brane is stable against small fluctuations of the metric in the gravity sector of the brane. 

In this context, we have explored a special case of the first order equations, in which one field may be investigated independently, changing the first order equation that describes the other field. We have then studied four distinct models, the first one engendering a surprising feature: the warp factor presents a sharp peak at the origin, which remind us of the thin brane behavior, and a smooth behavior outside the origin, as a usual thick brane. This induces a divergence in the energy density, at $y=0$, but this divergence is integrable and the total energy is still zero. In the second model, we have taken specific scalar field configurations that engender a parameter which is useful to tune an internal structure to the brane. In the third model, we have used the novel kinklike configuration described before in Ref. \cite{multikink} to create a brane with multiple internal structures, in which a parameter adjusts their distances to the center of the brane. Finally, in the fourth model we added a constant, which introduced an asymmetry to the brane, in addition to the multiple internal structure already shown in the third model. The asymmetry also changed the cosmological constant, giving rise to different scenarios, with distinct asymptotic geometries. 

An interesting perspective could be the addition of boson, fermion and gauge fields, to see how the models are able to localize those fields inside the brane; see, e.g., Refs. \cite{fl1,fl2,gl0,gl1} and references therein. Another issue concerns the use of the models studied in the present work to describe specific issues, such as the domain-wall/brane-cosmology correspondence \cite{DW1,DW2}, black hole formation \cite{BH} and modified models in the presence of Lagrange multiplier \cite{LM}. An issue of current interest concerns the new braneworld scenario developed in \cite{LM}, and we are now investigating the possibility to trade the second field $\chi$ of the model \eqref{lagran} by the Lagrange multiplier of the formulation developed in \cite{LM}. Another line of research related to modifications of the gravity sector of the braneworld model, concerns changing $R$ to $F(R)$, or to the Born-Infeld or Gauss-Bonnet or other generalized possibilities, as the ones described for instance in Refs. \cite{fr1,fr2,fr3,fr4,fr5}. Other issues concern the building of a Hamilton-Jacobi description \cite{HJ} for the braneworld model developed in the present work and, in the case of the asymmetric brane, the study of cosmic evolution, following the lines of Refs. \cite{N1,N2,N3}. We are now investigating some possibilities, hoping to report on them in the near future.\\

\acknowledgements{The work is supported by the Brazilian agencies Coordena\c{c}\~ao de Aperfei\c{c}oamento de Pessoal de N\'ivel Superior (CAPES), grants Nos. 88882.440250/2019-01 (DAF) and 88887.463746/2019-00 (MAM), Conselho Nacional de Desenvolvimento Cient\'\i fico e Tecnol\'ogico (CNPq), grants Nos. 303469/2019-6 (DB) and 404913/2018-0 (DB), and by Para\'iba State Research Foundation (FAPESQ-PB), grant No. 0015/2019.}


\end{document}